\documentclass[preprint2]{aastex}

\usepackage{bm, graphicx}

\received{2006 December 1}
\accepted{2007 January 10}

\slugcomment{In press, ApJ---est. {\bf 660}, No. 1, 2007 May 1}

\shorttitle{GENERALIZED SQUASHING FACTORS}
\shortauthors{TITOV}

\newcommand{\Rc}{R_{\rm c}}
\newcommand{\ee}{{\bm \epsilon}}
 
\begin{document}

\title{GENERALIZED SQUASHING FACTORS \\ FOR COVARIANT DESCRIPTION 
         OF \\ MAGNETIC CONNECTIVITY IN THE SOLAR CORONA} 


\author{V. S. Titov}
\affil{SAIC, 10260 Campus Point Drive, San Diego, CA 92121-1578} 
\email{titovv@saic.com
}





\begin{abstract}
The study of magnetic connectivity in the solar corona reveals
a need to generalize the field line mapping technique to arbitrary
geometry of the boundaries and systems of coordinates. Indeed, the
global description of the connectivity in the corona requires the
use of the photospheric and solar wind boundaries. Both are closed
surfaces and therefore do not admit a global regular system of
coordinates. At least two overlapping regular systems of coordinates
for each of the boundaries are necessary in this case to
avoid spherical-pole-like singularities in the coordinates of the
footpoints. This implies that the basic characteristic of magnetic
connectivity---the squashing degree or factor $Q$ of elemental flux
tubes \citep{Titov2002}---must be rewritten in covariant form.
Such a covariant expression of $Q$ is derived in this work. The
derived expression is very flexible and highly efficient for
describing the global magnetic connectivity in the solar corona.
In addition, a general expression for a new characteristic $Q_\perp$
which defines a squashing of the flux tubes in the directions
perpendicular to the field lines is determined. This new quantity
makes it possible to filter out the quasi-separatrix layers
whose large values of $Q$ are caused by a projection effect at the
field lines nearly touching the photosphere.
Thus, the value $Q_\perp$ provides a much more precise description
of the volumetric properties of the magnetic field structure.  The
difference between $Q$ and $Q_\perp$ is illustrated by comparing their
distributions for two configurations, one of which is the
Titov-D\'{e}moulin (1999) model of a twisted magnetic field.
\end{abstract}



\section{INTRODUCTION}
	\label{s:intro}

The structure of magnetic field is often an important factor in many energetic processes in the solar corona.
This especially refers to the topological features of magnetic structure such as null points, separatrix surfaces, and separator field lines.
They serve as preferred sites for the formation of current sheets and the corresponding accumulation of the free magnetic energy \citep{Sweet1969, Baum1980, Syrovatskii1981, Lau1990, Longcope1996, Priest1996, Priest2000, Longcope2001}.
The magnetic reconnection process induced in the current sheets at some critical parameters allows the accumulated magnetic energy to convert into other forms: thermal, radiative and kinetic energy of plasma and accelerated particles.
This process is considered to be a driving mechanism of many energetic phenomena in the solar atmosphere  \citep{Priest2000, Parker1979, Parker1994}.

Over the last decade, it also became clear that the geometrical analogs of the separatrices \citep{Longcope1994a, Longcope1994b, Titov1999a, Titov2002a, Titov2002}, the so-called quasi-separatrix layers (QSLs, \citep{Priest1995, Demoulin1996a, Demoulin1997}), have similar properties.
There are indications that the QSLs are probably more ubiquitous than the true separatrices \citep{Titov2002}.
This increases the significance of the problem of determining QSLs in a given magnetic configuration.
In comparison with the separatrices, the determining of QSLs requires a more sophisticated technique, which is based on a point-wise analysis of the magnetic field line connectivity.
The basic quantity in this technique is the squashing degree or factor $Q$ of elemental magnetic flux tubes.
This quantity has previously been defined for the planar geometry \citep{Titov1999a, Titov2002a, Titov2002}, which provides a good approximation for describing magnetic structures in active regions with the characteristic size smaller than the solar radius $R_{\sun}$.

Such an approximation, however, is hardly applicable for a global description of magnetic connectivity in the solar corona including the open magnetic field of the coronal holes.
The corresponding large-scale structure of magnetic fields is also of substantial interest for solar physics, especially, for understanding solar eruptions.
So the respective generalization of the above technique must allow us to determine $Q$ for the coronal volume bounded by the photospheric and solar-wind surfaces.
This immediately raises technical problems, which do not exist in the case of the planar geometry.
First, both these boundary surfaces are closed, and therefore, none of them admits a global regular system of coordinates.
To avoid a coordinate singularity of a spherical-pole type, at least two overlapping coordinate systems (coordinate charts) must be used in this case for describing the locations of the field line footpoints on each of the boundries.
Second, the solar-wind boundary surface generally cannot be a sphere, but some other curvilinear surface whose geometry depends on the coronal magnetic field \citep{Levin1982}.
These two requirements of the technique can be satisfied only by using a covariant approach to the description of $Q$ with the coordinate systems that are generally different for each of the boundaries.
The derivation of such a covariant expression for the squashing factor is one of the goals of the present work.

The second goal of the work is to make an essential refinement of the squashing factor itself.
The problem is that the large values of $Q$ may be caused not only by squashing of elemental flux tubes in the volume but also by a projection effect at the boundary surfaces.
The latter occurs at the field lines which are nearly touching the boundary at least at one of the footpoints.
This effect, in particular, takes place in the vicinity of the bald patches (BPs, \citep{Titov1993}), which are the segments of the photospheric polarity inversion line (PIL), where the coronal field lines touch the photosphere.
When analyzing magnetic connectivity, it is important to discriminate between the projection effect and volumetric squashing.
For this purpose, we derive a covariant expression for the perpendicular squashing factor $Q_\perp$, which describes the squashing of elemental flux tubes only in the directions orthogonal to the field lines.

Sections \ref{s:Q_cov} and \ref{s:Q_perp} present the derivations of $Q$ and $Q_{\perp}$ and demonstrate on the examples of planar and spherical geometry how to apply these general expressions.
The difference between $Q$ and $Q_{\perp}$ is considered in detail in \S             \ref{s:QvsQp} by calculating and comparing these quantities for two particular magnetic configurations.
The obtained results are summarized in \S \ref{s:s}.


\section{COVARIANT FORM OF THE \\ SQUASHING FACTOR}
	\label{s:Q_cov}

\begin{figure}[htbp]
\epsscale{0.9}
\plotone{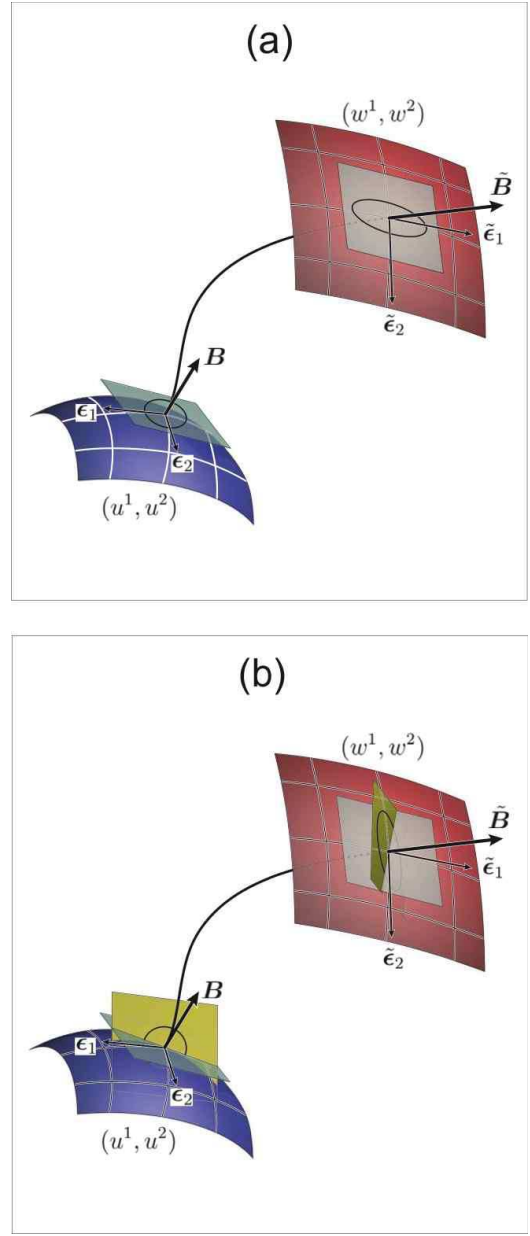}
\caption{Linearized field line mapping of a circle into an ellipse between tangent planes at the launch and target boundaries ({\it a}) and between the planes perpendicular to the field line at its footpoints ({\it b}).  In general, different and arbitrary coordinates $(u^{1},u^{2})$ and $(w^{1},w^{2})$ with covariant bases $(\ee_{1}, \ee_{2})$ and $(\tilde{\ee}_{1}, \tilde{\ee}_{2})$, respectively, are assumed at the launch and target boundaries.
	\label{f:f1} }
\end{figure}

Consider a plasma-magnetic configuration in a finite volume with a smooth boundary of an arbitrary shape.
It may generally consist of two or even more surfaces,---for example, the photosphere and the solar-wind surface form a boundary for the entire solar corona.
Each of the two footpoints of a given magnetic field line may belong in general to any of these surfaces.
We will use the designations ``launch" and ``target" for the footpoints and parts of the boundary surfaces at which the field lines start and end up.
Let $(u^{1},u^{2})$ and $(w^{1},w^{2})$ be the systems of curvilinear coordinates at the launch and target boundaries, respectively.
The magnetic field lines connecting these boundaries define a mapping $(u^{1},u^{2}) \rightarrow (w^{1},w^{2})$ determined by some vector-function $(W^{1}(u^{1},u^{2}), W^{2}(u^{1},u^{2}))$.
The local properties of this mapping are described by the Jacobian matrix
 \begin{eqnarray}
  D = \left[
         \frac{\partial W^{i}}{\partial u^{j}} 
      \right] .
  \label{D}
 \end{eqnarray}
For each field line, this matrix determines a linear mapping from the tangent plane at the launch footpoint to the tangent plane at the target footpoint, so that a circle in the first plane is mapped into an ellipse in the second plane (Fig. \ref{f:f1}a).
The aspect ratio of such an ellipse defines the degree of a local squashing  of elemental flux tubes, which means that any infinitesimal circle centered at a given launch point is mapped along the field lines into an infinitesimal ellipse with this aspect ratio at the target footpoint.
This generalizes a coordinate-free definition of the squashing factor to the case of curvilinear boundaries, whose tangent planes are generally not the same, as is in the case of plane boundaries considered in \citet{Titov1999a, Titov2002a} and \citet{Titov2002}.

%

To derive an analytical expression for the aspect ratio of the above ellipse, let us introduce first a vector function ${\bm R}(u^{1},u^{2})$ that describes in a three-dimensional (3D) Cartesian system of coordinates the locations of the footpoints at the launch boundary.
Then the vectors
\begin{eqnarray}
  \ee_{k} = \frac{\partial {\bm R}}{\partial u^{k}}, \quad k=1,2,
	\label{Ek}
\end{eqnarray}
determine at this boundary the covariant vector basis tangent to the $u$-coordinate lines.
Thus,
\begin{eqnarray}
	g_{lk} = \ee_{l} {\bm \cdot} \ee_{k}, \quad l,k=1,2,
	\label{glk}
\end{eqnarray}
is the corresponding covariant metric tensor, which determines local lengths and angles at the launch boundary.
The dot here stands for the usual scalar product in 3D Euclidean space.

Using (\ref{Ek}) and the standard Gramm-Schmidt procedure, one can construct an orthonormal basis
\begin{eqnarray}
  \bm{e}_{1} &=& \frac{\ee_{1}}{\sqrt{g_{11}}} ,
	\label{e1}
  \\
  \bm{e}_{2} &=& \frac{g_{12}}{\sqrt{g\, g_{11}}} \ee_{1}  
                 -\sqrt{\frac{g_{11}}{g}} \ee_{2} .
	\label{e2}
\end{eqnarray}
Hereafter $g \equiv \det \left[g_{lk}\right]$ is a determinant of the {\it covariant} metric tensor.
Now, any point of a circle of unit radius in the plane tangent to the launch boundary is represented by the vector
\begin{eqnarray}
	{\bm o} = \cos\vartheta\, {\bm e}_{1} + \sin\vartheta\, {\bm e}_{2},
	\label{o}
\end{eqnarray}
whose angle parameter $\vartheta \in [0,2\pi)$.

Suppose that the vector-function $\tilde{\bm R}(w^{1},w^{2})$ defines the points at the target boundary, then
\begin{eqnarray}
	\tilde{\bm o} = o^{k} \frac{\partial W^{i} }{\partial u^{k}}\tilde{\ee}_{i} 
	\label{ot} 
\end{eqnarray}
is the field-line mapping image of ${\bm o}$ at the tangent plane of the target boundary, where the corresponding covariant basis vectors, parallel to the $w$-coordinate lines, are
\begin{eqnarray}
  \tilde{\ee}_{i}=\frac{\partial \tilde{\bm R}}{\partial w^{i}} .
	\label{Ei}
\end{eqnarray}
Hereafter a summation over repeating indices with their values running from 1 to 2 is assumed.

With varying $\vartheta$, the vector $\tilde{\bm o}$ traces in this
plane an ellipse such that
\begin{eqnarray}
 	\tilde{\bm o}^{2} &\equiv& 
	g^{*}_{ij} \tilde{o}^{i} \tilde{o}^{j} 
	= \frac{1}{2}	g^{*}_{ij} \frac{\partial W^{i}}{\partial u^{k}}
			   \frac{\partial W^{j}}{\partial u^{l}}
			   	         \nonumber \\
		&& \times \left[ e_{1}^{k} e_{1}^{l} + e_{2}^{k} e_{2}^{l}+
		       \cos 2\vartheta 
		       \left( e_{1}^{k} e_{1}^{l} -
					      e_{2}^{k} e_{2}^{l} \right)
		       \right.  \nonumber \\
		        && \left. 
 		      \qquad +\sin 2\vartheta \left( e_{1}^{k} e_{2}^{l} +
					      e_{1}^{l} e_{2}^{k} \right)
		\right] , 
	\label{ot2}
\end{eqnarray}
where the asterisk indicates that $g^{*}_{ij}(u^{1},u^{2})$ is a result of evaluating $\tilde{g}_{ij}(w^{1}, w^{2})$ at the target footpoint $(w^{1}, w^{2}) = (W^{1}(u^{1},u^{2}), W^{2}(u^{1},u^{2}))$.

After some simple trigonometry and lengthy algebra using
equations (\ref{Ek})--(\ref{e2}), equation (\ref{ot2}) is reduced to
\begin{eqnarray}
	\tilde{\bm o}^{2} = \frac{1}{2} 
		(N^{2}+\sqrt{N^{4}-4\Delta^{2}} \sin
		2\tilde{\vartheta}), 
	\label{ot2m} 
\end{eqnarray}
where $N^{2}$ and $\Delta$ are determined by 
\begin{eqnarray}
	N^{2} & = & \frac{\partial W^{i}}{\partial u^{k}} g^{*}_{ij}
			              \frac{\partial W^{j}}{\partial u^{l}} g^{lk},
	\label{N2c}
\\
	\Delta & = & \sqrt{\frac{g^{*}}{g}} 
		\frac{\partial(W^{1},W^{2})}{\partial(u^{1},u^{2})},
	\label{Dltc}
\end{eqnarray}
in which $g$ and $g^{*}$ denote the determinants of {\it covariant} metric tensors at the launch and target footpoints, respectively.
The components of the {\it contravariant} metric tensor $g^{lk}$ can be viewed here as elements of the inverted matrix $\left[g_{lk}\right]^{-1}$ of the covariant metric. 
The value $\tilde{\vartheta}$ is simply $\vartheta$ plus an additional value, which is independent of $\vartheta$ and whose expression does not matter for the present consideration.

What actually matters is that $\sin 2\tilde{\vartheta}$ runs values from $-1$ to $+1$ when ${\bm o}(\vartheta)$ and $\tilde{\bm o}(\vartheta)$ are tracing, respectively, the above circle and ellipse.
The minimum $-1$ and maximum $+1$ correspond here to the minor and major axes of the ellipse, respectively, so that its aspect ratio is
\begin{eqnarray}
	\frac{\tilde{o}_{\rm max}}{\tilde{o}_{\rm min}} &=&
		 \left( \frac{N^{2}+\sqrt{N^{4}-4\Delta^{2}}}
			     {N^{2}-\sqrt{N^{4}-4\Delta^{2}}}
		 \right)^{1/2} 
		 \nonumber \\
		 &=&
	 \frac{N^{2}}{2|\Delta|} +
	 \sqrt{ \left(  \frac{N^{2}}{2|\Delta|} \right)^{2}-1} .
	\label{are}
\end{eqnarray}
The large values of this ratio do not differ much from its asymptotic
value
\begin{eqnarray}
    Q = N^{2}/|\Delta| .
	\label{Q}
\end{eqnarray}
Note also that $Q\ge 2$, since inverting equation (\ref{are}) yields $Q= \tilde{o}_{\rm max}/\tilde{o}_{\rm min} + \tilde{o}_{\rm min}/\tilde{o}_{\rm max}$ and $\tilde{o}_{\rm max}/\tilde{o}_{\rm min} \ge 1$.
Therefore equation (\ref{Q}) will be used as a covariant definition of the squashing factor.

It is evident from the derivation of $Q$ that this value is invariant to the direction of field line mapping.
Indeed, the inverse mapping implies locally that $1/\tilde{o}_{\rm min}$ is a maximum stretching coefficient and $1/\tilde{o}_{\rm max}$ is a minimum shrinking coefficient.
Such coefficients will coincide with the lengths of the major and minor axes of the ellipse obtained from a circle of a unit radius due to this inverse mapping.
Thus, although this new ellipse has different lengths of axes, their ratio is the same as for the previous one, which proves the statement.
A formal proof of the statement is also not difficult to obtain by using the derived expressions of $N^{2}$ and $\Delta$ in a similar way as in the case of plane boundaries \citep{Titov1999a}.
The invariancy of $Q$ to the direction of field line mapping justifies its status of a correct measure for the magnetic connectivity.

Note also that $\Delta$ for a given infinitesimal flux tube is a ratio of its cross section areas at the target and launch points.
Therefore, since the magnetic flux is conserved along the tubes, this value coincides with the corresponding inverse ratio of the normal field components, so that 
\begin{eqnarray}
	\Delta = B_{n}/B^{*}_{n} ,
	\label{Dlt2}
\end{eqnarray}
where $B_{n}$ and $B^{*}_{n}$ are normal components of the magnetic field to the boundaries at the conjugate launch and target footpoints.
In practice, the numerical calculation of $\Delta$ through this ratio is more precise than that given by equation (\ref{Dltc}) and therefore it should be used for computing $\Delta$ in equation (\ref{Q}).

The above mathematical construction is related to the Cauchy-Green deformation tensor \citep{Marsden2002} known in the theory of elasticity.
It can be written in our notations as
\begin{eqnarray}
  C_{kl} = \frac{\partial W^{i}}{\partial u^{k}} g^{*}_{ij}
			              \frac{\partial W^{j}}{\partial u^{l}} ,
	\label{C}
\end{eqnarray}
where $(W^{1}(u^{1},u^{2}), W^{2}(u^{1},u^{2}))$ and $g_{ij}$ represent, respectively, a finite deformation and covariant metric tensor of an elastic two-dimensional body.
The contraction of the Cauchy-Green tensor with a pair of orthonormal vectors ${\bm e}_{m}$ and ${\bm e}_{n}$ yields the tensor
\begin{eqnarray}
  \tilde{C}_{mn} = C_{kl} e^{k}_{m} e^{l}_{n}
\end{eqnarray}
such that its eigenvalues coincide with the squared semiaxes $\tilde{o}^{2}_{\rm max}$ and $\tilde{o}^{2}_{\rm min}$ of the above ellipse.
The square root of their ratio defines in accordance with (\ref{are}) and (\ref{Q}) the squashing factor $Q$.

It should be emphasized that this analogy is possible only in our general approach, where two independent systems of coordinates are used for describing the location of the conjugate footpoints.
This allows us to apply coordinate transformations only at the launch boundary, while keeping the coordinates at the target boundary unchanged.
With respect to these transformations, the object defined by (\ref{C}) does behave as a covariant second-rank tensor.
The latter is not valid, however, if one global 3D system of coordinates is used for describing the entire field configuration and so both boundaries are subject then to coordinate transformations.

This has only a methodological meaning and does not exclude, of course, an application of the derived expressions to such particular cases.
For example, consider a closed magnetic configuration in the half space $x^{3}\ge0$ with the global Cartesian coordinates $(x^{1},x^{2},x^{3})\equiv(u^{1},u^{2},x^{3})\equiv(w^{1},w^{2},x^{3})$ and the photospheric boundary plane $x^{3}=0$.
The field line mapping is then given by $\left( X^{1}(x^{1},x^{2}), X^{2}(x^{1},x^{2}) \right)$.
There are no more differences between upper and low indices and contravariant $g^{lk}$ and covariant $g_{ij}^{*}$ metrics; the latter simply turn into Kronecker symbols $\delta^{kl}$ and $\delta_{ij}$.
So equations (\ref{N2c}), (\ref{Dltc}) and (\ref{Dlt2}) are reduced to
\begin{eqnarray}
  N^{2} &=&
    \frac{\partial X^{i}}{\partial x^{k}}
    \frac{\partial X^{i}}{\partial x^{k}}, 
	\label{N2} \\
	\Delta &=&
	  \frac{\partial (X^{1},X^{2})}{\partial (x^{1},x^{2})}
	  = \frac{B_{3}}{B^{*}_{3}} , 
	\label{Dlt} 
\end{eqnarray}
as required in this case \citep{Titov2002}.

Consider now a more complicated class of configurations, where both open and closed magnetic field lines are present.
Let the configuration be described in one global system of coordinates $(r,\theta,\phi)$, where $r=R_{\sun}$ corresponds to the photospheric launch boundary, while $r=R_{*}$ represents the target boundary.
For the open field lines reaching the spherical solar-wind boundary of radius $R_{\rm SW}$, we put $R_{*}=R_{\rm SW}$, while for the closed ones we take $R_{*}=R_{\sun}$.
Thus, $u^{1}=w^{1}=\phi$, $u^{2}=w^{2}=\theta$ and the field line mapping is $(\Phi(\phi, \theta),\Theta(\phi, \theta))$, which yields
\begin{eqnarray}
	[ g^{*}_{ij} ] = 
	\left( \begin{array}{cc}
		R_{*}^{2}\sin^{2}\Theta 	&0 	  \\
		0				&R_{*}^{2}
	       \end{array}
	\right)
	\label{gijs*} 
\end{eqnarray}
and
\begin{eqnarray}
	[ g^{lk} ] = 
	\left( \begin{array}{cc}
		R_{\sun}^{-2}\sin^{-2}\theta 	&0 	\\
		0				&R_{\sun}^{-2}
	       \end{array}
	\right) ,
	\label{gkls} 
\end{eqnarray}
where the contravariant metric $g^{lk}$ at the launch boundary is obtained from the corresponding covariant metric by inverting it simply as a $2\times2$ matrix.
Using these expressions and equation (\ref{N2c}), we obtain
\begin{eqnarray}
  N^{2} &=& \frac{R^{2}_{*}}{R^{2}_{\sun}}
  \left[
    \left( \frac{\sin\Theta}{\sin\theta} 
		       \frac{\partial \Phi }{\partial \phi }
     \right)^{2} +
    \left(
       \sin\Theta \frac{\partial \Phi }{\partial
			\theta}
    \right)^{2} \right. \nonumber \\ 
    && \left. +
    \left(
       \frac{1}{\sin\theta} \frac{\partial \Theta }{\partial \phi }
    \right)^{2} +
    \left(
       \frac{\partial \Theta }{\partial \theta }
    \right)^{2} 
  \right] .
	\label{N2sph}
\end{eqnarray}
Equations (\ref{Dltc}) and (\ref{Dlt2}) in this case become
\begin{eqnarray}
  \Delta = \frac{R^{2}_{*}}{R^{2}_{\sun}}
    \frac{\sin\Theta}{\sin\theta}\,
    \frac{\partial (\Phi,\Theta)}{\partial (\phi,\theta)} 
   =\frac{B_{r}}{B^{*}_{r}} .
	\label{Dltsph}
\end{eqnarray}

The obtained expressions for $N^{2}$ and $\Delta$ have seeming singularities at the poles, where they actually reduce in the generic case to resolved indeterminacies with $\Phi(\phi,\theta)$ and $\Theta(\phi,\theta)$ proportional to $\sin \theta$.
These indeterminacies are artificial and unrelated to some special properties of the magnetic
structure; they are caused by the pole singularities inherent in the used global spherical system of coordinates.
Moreover, their appearance is unavoidable in any other global system of coordinates on a closed sphere-like surface because of its intrinsic topological properties.
Therefore this may generally reduce precision of a numerical evaluation of $Q$ near the pole-like points.
To avoid such indeterminacies, at least two overlapping coordinate charts on each of the spherical boundaries are required.
For this purpose, it is sufficient to use two systems of spherical coordinates turned with respect to each other on the right angle in the $\theta$-direction.
Switching from one of such systems in its polar regions to the other, as suggested, for example, by \citet{Kageyama2004}, makes it possible to resolve the problem of the pole indeterminacies.
The required expressions for such calculations of $Q$ can be obtained again from equations (\ref{N2c}) and (\ref{Dltc}) with properly modified metric tensors.


\section{PERPENDICULAR COVARIANT \\ SQUASHING FACTOR}
	\label{s:Q_perp}

To find the perpendicular squashing factor $Q_{\perp}$, we need to know the field line mapping between infinitesimal planes orthogonal to the field lines at the conjugate footpoints (Fig. \ref{f:f1}b).
Note first that the projection effect at the boundaries is local, because it depends only on the orientations of the tangent planes at the boundaries with respect to the vectors of magnetic field at the footpoints.
So the required mapping between the indicated orthogonal planes can be obtained from the respective mapping between the tangent planes by correcting it only at such footpoints.
This implies that $Q_{\perp}$ can be expressed in terms of the same values as $Q$ and, in addition, the field vectors ${\bm B}$ and $\tilde{\bm B}$, respectively, at the launch and target footpoints. 

We will derive $Q_{\perp}$ by using the same procedure as for $Q$ while modifying it in accordance with the above comments.
The vector ${\bm o}$ tracing the circle of unit radius is given by the same expression (\ref{o}).
However, since it lies now in the plane perpendicular to the field line at the launch point, the corresponding orthonormal basis is chosen to be orthogonal to ${\bm B}$, so that
\begin{eqnarray}
  {\bm e}_{1} & = & \frac{{\bm B}{\bm \times} \ee_{1}}
    {\left|{\bm B}{\bm \times} \ee_{1}\right|} ,
	\label{e1p}
  \\
  {\bm e}_{2} & = & \frac{{\bm B}{\bm \times} {\bm e}_{1}}
    {\left|{\bm B}{\bm \times} {\bm e}_{1}\right|} .
	\label{e2p}
\end{eqnarray}
This vector ${\bm o}$ is mapped along the field lines into a vector $\tilde{\bm o}$ lying in the plane perpendicular to the local field $\tilde{\bm B}$ at the target footpoint.
The respective mapping ${\rm d}W^{\perp}$ can be represented as a composition $P^{-1} \circ {\rm d}W \circ P$ of three others according to the following diagram:
\begin{equation}
\begin{array}{ccc}
{\bm o} & {{\scriptscriptstyle{\rm d}W^{\perp}} \atop \longrightarrow} & \tilde{\bm o}\\
\hspace{0.5 em}\downarrow \scriptscriptstyle{P} && \hspace{1.3 em}\uparrow \scriptscriptstyle{P}^{-1}\\
\underline{\bm o} & {{\scriptscriptstyle{\rm d}W} \atop \longrightarrow} & \underline{\tilde{\bm o}}
\end{array}
\end{equation}
Here the mapping $P$ projects the vector ${\bm o}$ along ${\bm B}$ onto the plane tangential to the launch boundary to yield the vector
\begin{eqnarray}
   \underline{\bm o} & = & {\bm o} - 
     \frac{\left(\bm{o\cdot}\ee_{1}{\bm \times} \ee_{2}\right)}
        {\left(\bm{B\cdot}\ee_{1}{\bm \times} \ee_{2}\right)} {\bm B} ,
	\label{o_}
\end{eqnarray}
which has a vanishing component along the vector $\ee_{1}{\bm \times} \ee_{2}$ perpendicular to such a plane.
Then this vector $\underline{\bm o}$ is mapped by the differential of the field line mapping ${\rm d}W$ determined by the Jacobian matrix (\ref{D}) into the vector
\begin{eqnarray}
   \underline{\tilde{\bm o}} & = & \underline{o}^{k} 
      \frac{\partial W^{i}}{\partial u^{k}} \tilde{\ee}_{i}
	\label{o_t}
\end{eqnarray}
which lies in the plane tangential to the target boundary.
Finally, the obtained vector $\underline{\tilde{\bm o}}$ is projected by $P^{-1}$ along $\tilde{\bm B}$ at the target footpoint onto the plane perpendicular to $\tilde{\bm B}$ to result in
\begin{eqnarray}
  \tilde{\bm o} & = & \underline{\tilde{\bm o}} - 
    \frac{\bigl(\tilde{\bm B}{\bm \cdot}\underline{\tilde{\bm o}}\bigr)}
         {\tilde{\bm B}^{2}}{\tilde{\bm B}} .
	\label{otp}
\end{eqnarray}

Eliminating now $\underline{\tilde{\bm o}}$ and $\underline{\bm o}$ from (\ref{o_})--(\ref{otp}) and using (\ref{o}) with the basis from equations (\ref{e1p})--(\ref{e2p}), we express $\tilde{\bm o}$ in terms of $\ee$-basis and magnetic field vectors at the conjugate footpoints.
This allows us to calculate $\tilde{\bm o}^{2}$ and then $Q_{\perp}$ in a similar way as done before when deriving $Q$.
The result is
\begin{eqnarray}
  Q_{\perp} & = & N^{2}_{\perp} /  |\Delta_{\perp}| ,
	\label{Qpc}
\\
  N_{\perp}^{2} & = &
	  \frac{\partial W^{i}}{\partial u^{k}} g^{*}_{\perp ij}
		\frac{\partial W^{j}}{\partial u^{l}}  g^{\perp lk} ,
	\label{N2pc}
\\
	\Delta_{\perp} & = & \sqrt{\frac{g^{*}_{\perp}}{g_{\perp}}} 
		\frac{\partial(W^{1},W^{2})}{\partial(u^{1},u^{2})} ,
	\label{Dltpc}
\end{eqnarray}
where the asterisk has the same meaning as in equations (\ref{ot2}), (\ref{N2c}) and (\ref{Dltc}).
Thus, the obtained $Q_{\perp}$ differs from $Q$ only in the form of the metric tensors, which are determined now by
\begin{eqnarray}
  g_{\perp ij}^{*} &= & \left(g_{ij} - \frac{B_{i}B_{j}} {\bm{B}^{2}}\right)^{*} ,
	\label{gpij*}
  \\
  g^{\perp lk} & = & g^{lk} + 
      \frac{g B^{l} B^{k}} {\left(\bm{B\cdot}\ee_{1}\bm{\times}\ee_{2}\right)^{2}} .
	\label{gplk}
\end{eqnarray}
The asterisk here implies automatically that the corresponding values refer to the target footpoint, therefore the tilde used for indicating this fact in intermediate equation (\ref{otp}) is omitted in the final expression (\ref{gpij*}).
One can also check that $\left[g^{\perp lk}\right] = \left[g_{\perp lk}\right]^{-1}$ and
\begin{eqnarray}
  g_{\perp} \equiv \det \left[ g_{\perp lk} \right] = 
     \frac{\left(\bm{B\cdot}\ee_{1} \bm{\times}\ee_{2}\right)^{2}}
          {\bm{B}^{2}} ,
	\label{gp}
\end{eqnarray}
so that equation (\ref{Dltpc}) reduces to
\begin{eqnarray}
  |\Delta_{\perp}| = \frac{|\bm{B}|}{|\bm{B}^{*}|}  
	\label{Dltpc2}
\end{eqnarray}
if, in addition, the magnetic flux conservation is taken into account.
This expression should be used instead of (\ref{Dltpc}) for computing $|\Delta_{\perp}|$ in (\ref{Qpc}) for the same reason that (\ref{Dlt2}) should be used in (\ref{Q}).
Thus, formulas (\ref{Qpc})--(\ref{Dltpc2}) completely define the covariant expression for $Q_{\perp}$.

Let us see now how these formulas work in the case of a closed magnetic configuration described in a global Cartesian system of coordinates $(x^{1},x^{2},x^{3}) \equiv (u^{1},u^{2},x^{3}) \equiv (w^{1},w^{2},x^{3})$ with the photospheric boundary plane $x^{3}=0$.
The formulas are significantly simplified in this case to yield 
\begin{eqnarray}
  N_{\perp}^{2} & = &
	  \frac{\partial X^{i}}{\partial x^{k}} \left(\delta_{ij} -
	             \frac{B^{*}_{i}B^{*}_{j}}{\bm{B}^{*2}}\right)
	                 \nonumber \\
	                 && \times
		\frac{\partial X^{j}}{\partial x^{l}} \left(\delta^{lk} + 
      \frac{B^{l} B^{k}}{\left(B_{3}\right)^{2}}\right),
	\label{N2pp}
\end{eqnarray}
where the values of the corresponding covariant and contravariant components of vectors and tensors do not differ from each other.

Note also that expression (\ref{N2pp}) apparently diverges near the PIL, where the normal field component $B_{3}$ vanishes.
This is actually not a true singularity but rather an indeterminacy, which is resolved to give a low limit of $Q_{\perp}\big|_{\rm PIL}=2$ if the PIL has no BPs.
At the BPs, such an indeterminacy is also resolved but it may generally have different limits of $Q_{\perp}>2$ at the left and right sides of BPs.
Thus, $Q_{\perp}$ may experience a jump when crossing BPs, unless the configuration is symmetric as in the example of \S \ref{s:HFTtwst}.

Similar to \S \ref{s:Q_cov}, consider also a more general class of configurations, where both open and closed field lines are present and bounded by a spherical solar-wind surface of radius $R_{\rm SW}$ and the photosphere of radius $R_{\sun}$ as before.
It is convenient in this case to use matrix notations, in which equation (\ref{N2pc}) is written as
\begin{eqnarray}
N^{2}_{\perp} = {\rm tr}\!\left(D^{\rm T} G^{*}_{\perp} 
	                                   D         G^{\perp} \right) ,
	\label{N2po}
\end{eqnarray}
where
\begin{eqnarray}
  D =  \left(
	   \begin{array}{cc}
      \displaystyle\frac{\partial \Phi}{\partial \phi} 
          &\displaystyle\frac{\partial \Phi}{\partial \theta}    \\ [8pt]
      \displaystyle\frac{\partial \Theta}{\partial \phi}
          &\displaystyle\frac{\partial \Theta}{\partial \theta}   
     \end{array} 
  \right) 
	\label{Do}
\end{eqnarray}
is the Jacobian matrix of the field line mapping.
The covariant and contravariant metrics at the target and launch boundaries, respectively, are determined by the following matrices:
\begin{eqnarray}
 && G^{*}_{\perp} \equiv \left[g_{\perp ij}\right]^{*} 
    \nonumber \\
  && =  R^{2}_{*} \left(
	   \begin{array}{cc} 
      \displaystyle \sin^{2}\Theta  \left(1-\frac{B_{\phi}^{*2}}{{\bm B}^{*2}}\right) 
        & \displaystyle -\sin\Theta \frac{B_{\phi}^{*} B_{\theta}^{*}}{\bm{B}^{*2}} \\ [8pt]
     \displaystyle -\sin\Theta\frac{B_{\phi}^{*} B_{\theta}^{*}}{\bm{B}^{*2}}
        & \displaystyle 1-\frac{B_{\theta}^{*2}}{\bm{B}^{*2}}   
     \end{array} \right) , \qquad
	\label{Gpo*}
	  \\ [8pt]
&&  G^{\perp} \equiv  \left[g^{\perp lk}\right]
     \nonumber \\
&&     =  R^{-2}_{\odot} \left(
	   \begin{array}{cc}
      \displaystyle\sin^{-2}\theta\left(1+\frac{B_{\phi}^{2}}{B_{r}^{2}}\right) 
          &\displaystyle \frac{B_{\phi}B_{\theta}}{\sin\theta\, B_{r}^{2}}    \\ [10pt]
      \displaystyle\frac{B_{\phi}B_{\theta}}{\sin\theta\, B_{r}^{2}}
          &\displaystyle 1+\frac{B_{\theta}^{2}}{B_{r}^{2}}   
     \end{array} \right) ,
	\label{Gpo}
\end{eqnarray}
where $R_{*}=R_{\sun}$ if a given footpoint belongs to a closed field line and $R_{*}=R_{\rm SW}$ otherwise.

The matrix $G^{\perp}$ has two types of singularities, which actually lead only to indeterminacies in expression (\ref{N2po}).
The first indeterminacies take place at the poles of spherical coordinates ($\theta=0,\pi$).
They are already discussed above in connection with equation (\ref{N2sph}). 
The second indeterminacies take place at the PIL, where $B_{r}=0$ .
They are resolved in much the same way as occurred in the previous case of Cartesian geometry to give a finite value of $Q_{\perp}\big|_{\rm PIL}$.
If the PIL has no BPs, the length of the field lines vanishes near the PIL, so that $D\rightarrow I$ and $G^{*}_{\perp} \rightarrow G^{\perp -1}$, which results in $N^{2}_{\perp} \rightarrow 2$.
The presence of BPs implies, however, a strong projection effect in their neighborhood and the corresponding singularities in the Jacobian matrix $D$ at the BPs.
These singularities are exactly of the same type and value as those in $G^{\perp}$ but opposite in signs.
So they cancel each other in equation (\ref{N2po}) to generally provide different limits at the different sides of the BPs, as discussed above for the case of the plane boundaries.

It should be also noted that there is one more type of plausible singularities not yet discussed.
These singularities have to appear if a given configuration has null points of magnetic field and the corresponding separatrix field lines.
In this case, the elemental flux tubes enclosing the separatrix field lines split at each of the null points to produce singularities in the derivatives of the Jacobian matrix.
Such singularities are due to volumetric rather than surface properties of magnetic configurations.
Therefore, if present, they appear in both $Q$ and $Q_{\perp}$ distributions thereby indicating the existence of magnetic nulls in the corona.
Since the numerical derivatives are estimated as a ratio of finite coordinate differences, their absolute values may not exceed the ratio of the coordinate range at the target boundary to a chosen increment of coordinates at the launch boundary.
The presence of this upper bound on possible values of the numerical derivatives prevents an overflow error in computations of the squashing factors at the footpoints of the null-point separatrices.
A more detailed consideration of this type of singularity requires a special study, which goes far beyond the scope of the present work.


\section{COMPARISON OF $Q$ and $Q_{\perp}$}
	\label{s:QvsQp}

To see the difference between $Q$ and $Q_{\perp}$, their distributions are compared below for several magnetic configurations. 
An emphasis is made on their potentiality for determining so-called hyperbolic flux tubes (HFTs).
They are defined as two intersecting QSLs with extended and narrow photospheric footprints characterized by very large values of $Q$.
Neglecting the possible curving and twisting of an HFT, its cross section variation in the longitudinal direction can be respresented as
\begin{equation}
			\includegraphics*[scale=0.25]{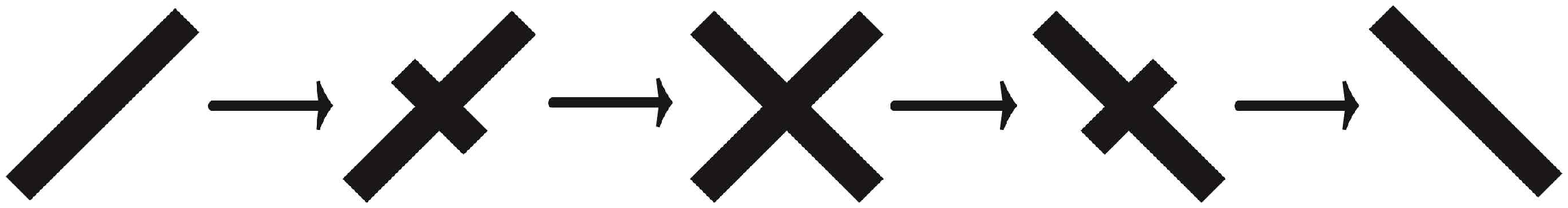}
	\label{-X-}
\end{equation}
In other words, the width of one of the QSLs is shrunken to the thickness of the other in the process of the mapping of their cross sections along magnetic field lines.
This is possible because the field lines in HFTs exponentially converge in one transversal direction and diverge in the other. 
Such a property is typical for hyperbolic flows in the theory of dynamical systems \citep{Arnold88}, which was coined in the term HFT \citep{Titov2002}.
The examples below demonstrate that $Q_{\perp}$ is a more accurate quantity than $Q$ for characterizing HFTs, although $Q$ also provides valuable complementary information on the magnetic structure.


\subsection{The simplest possible hyperbolic flux tube}
	\label{s:HFTsmpl}

The most simple magnetic configuration in which one would expect the presence of an HFT is the so-called X-line configuration, whose field ${\bm B}$ is determined in Cartesian coordinates $(x,y,z)$ by
\begin{eqnarray}
	{\bm B} = (-hx,hy,1) = {\bm \nabla}\xi {\bm \times} {\bm \nabla}\eta ,
	\label{BX}
\end{eqnarray}
where $h=\bar{h}L/B_{\|}$ is a dimensionless field gradient characterizing the strength of the transverse field $\bar{h}L$ compared to the longitudinal field $B_{\|}$ on a characteristic length scale $L$.
The right-hand side of (\ref{BX}) represents the same field in terms of the Euler potentials
\begin{eqnarray}
	\xi  &=& x\, e^{ hz} , 
	  \label{xi}  \\
  \eta &=& y\, e^{-hz} ,
   	\label{eta}
\end{eqnarray}
which are constant along the field lines.
The use of $\xi$ and $\eta$ significantly simplifies the calculation of the squashing factors in this configuration.
Suppose that its volume is restricted by $|z| \le 1$, so that the planes $z=\pm1$ are the corresponding boundaries.
The constancy of $\xi$ and $\eta$ means that the boundary points $(x_{-},y_{-})$ and $(x_{+},y_{+})$ are related by $x_{-} e^{-h} = x_{+} e^{h}$ and $y_{-} e^{h} = y_{+} e^{-h}$.
In terms of the notations used in expressions (\ref{N2}) for the plane boundaries, this means that $X_{1}=e^{2h} x_{1}$ and $X_{2} = e^{-2h} x_{2}$.
Since $B_{z}=1$ everywhere, equation (\ref{Dlt}) reduces simply to $|\Delta|=1$, so that the resulting squashing factor is
\begin{eqnarray}
	Q  = 2 \cosh(4h) .
	\label{QX}
\end{eqnarray}
Thus, $Q$ is constant over the entire planes $z=\pm1$ and, hence, it does not determine any QSL in the X-line configuration with the plane boundaries.
\citet{Priest1995} has arrived at the same conclusion by using the norm $N$ defined by (\ref{N2}), since $B_{z}=1$, $|\Delta|=1$ and so $N=\sqrt{Q}$ in this configuration.
However, they have found with the help of $N$ some evidences of the presence of QSLs in the X-line configuration, if it is bounded by cubic, hemispheroidal and spherical surfaces.

On the other hand, it is clear from the general point of view that there should be an HFT in such a configuration irrespective of the shape of the boundary surfaces.
Indeed, this configuration has in the unbounded space two genuine separatrix planes $x=0$ and $y=0$, which can be regarded as a limiting case of the bounded configuration with the boundaries moved off to infinity.
Taking into account also that $h \sim L$, one could expect that the proper measure for QSLs must be growing with $h$ near the planes $x=0$ and $y=0$ much stronger than in the remaining volume to indicate the corresponding QSLs near these planes.
This would provide in the limit of large $h$ an expected continuous transition of such bounded configurations with QSLs to the unbounded configuration with the genuine separatrices.
The failure of $Q$ in determining an HFT in this simple case looks very surprising in light of its remarkable success in other more complicated field configurations (see \S \ref{s:HFTtwst}).
The reason for this failure lies actually in the above-mentioned projection effect, which is extremely large for the chosen type of boundaries.
The transverse component here grows linearly with the distance from the X-line $x=y=0$, while the longitudinal component remains constant.
So the farther a given field line meets the boundary from the X-line, the more they become aligned with each other.

Such an explanation is fully confirmed by calculations of the perpendicular squashing factor $Q_{\perp}$ in this case.
To derive an analytical expression of $Q_{\perp}$, let us choose $(\xi,\eta)$ as coordinates on both boundary surfaces, so that $u^{1}=w^{1}=\xi$ and $u^{2}=w^{2}=\eta$ in (\ref{Qpc})--(\ref{Dltpc2}).
Assume for generality that the boundaries are defined by $z=Z_{\pm}(\xi,\eta)$, then  according to (\ref{xi}) and (\ref{eta}) the vector functions
\begin{eqnarray}
  {\bm R}(\xi,\eta)       & = & (\xi\, e^{-hZ_{-}},\eta\, e^{hZ_{-}}, Z_{-}) ,
	\label{R}  \\
	\tilde{\bm R}(\xi,\eta) & = & (\xi\, e^{-hZ_{+}},\eta\, e^{hZ_{+}}, Z_{+}) 
	\label{Rt}
\end{eqnarray}
define, respectively, the launch and target boundary surfaces.
These formulas are needed for calculating (\ref{Ek}), (\ref{glk}), (\ref{Ei}), (\ref{N2pc}) (\ref{gpij*}), (\ref{gplk}), (\ref{Dltpc2}) and (\ref{BX}) in the chosen coordinates $(\xi,\eta)$.
With the help of such calculations, equation (\ref{Qpc}) yields
\begin{eqnarray}
	Q_{\perp}= \frac{2\cosh\left[2h(Z_{+}-Z_{-})\right] + 
	  h^{2} \left(X_{+}^{2}+Y_{+}^{2}+X_{-}^{2}+Y_{-}^{2}\right)}
	  {\sqrt{\left[1+h^{2}\left(X_{+}^{2}+Y_{+}^{2}\right)\right]
	   \left[1+h^{2}\left(X_{-}^{2}+Y_{-}^{2}\right)\right]}} ,
	\label{QpX}
\end{eqnarray}
in which
\begin{eqnarray}
	X_{\pm} &=& \xi \, e^{-hZ_{\pm}} , 
		\label{Xpm} \\
	Y_{\pm} &=& \eta\, e^{ hZ_{\pm}}  
	  \label{Ypm} 
\end{eqnarray}
determine $(x,y)$ coordinates of the conjugate footpoints at the defined boundaries.

These expressions determine $Q_{\perp}$ as a function of the Euler potentials $\xi$ and $\eta$.
Note also that $\xi=x$ and $\eta=y$ at $z=0$, so expressions (\ref{QpX})--(\ref{Ypm}) define in addition the distribution of $Q_{\perp}$ in the plane $z=0$, to which $Q_{\perp}$ is mapped along the field lines from the boundaries.
More generally, the combination of these expressions with equations (\ref{xi}) and (\ref{eta}), resolved with respect to $x$ and $y$ as
\begin{eqnarray}
	x &=& \xi \, e^{-hz} , 
	  \label{x}  \\
  y &=& \eta\, e^{ hz} ,
   	\label{y}
\end{eqnarray}
provides a parametrical representation of $Q_{\perp}$, with $\xi$ and $\eta$ as parameters, in any plane $z={\rm const}$ between the boundaries.

\begin{figure}[htbp]
\epsscale{0.8}
\plotone{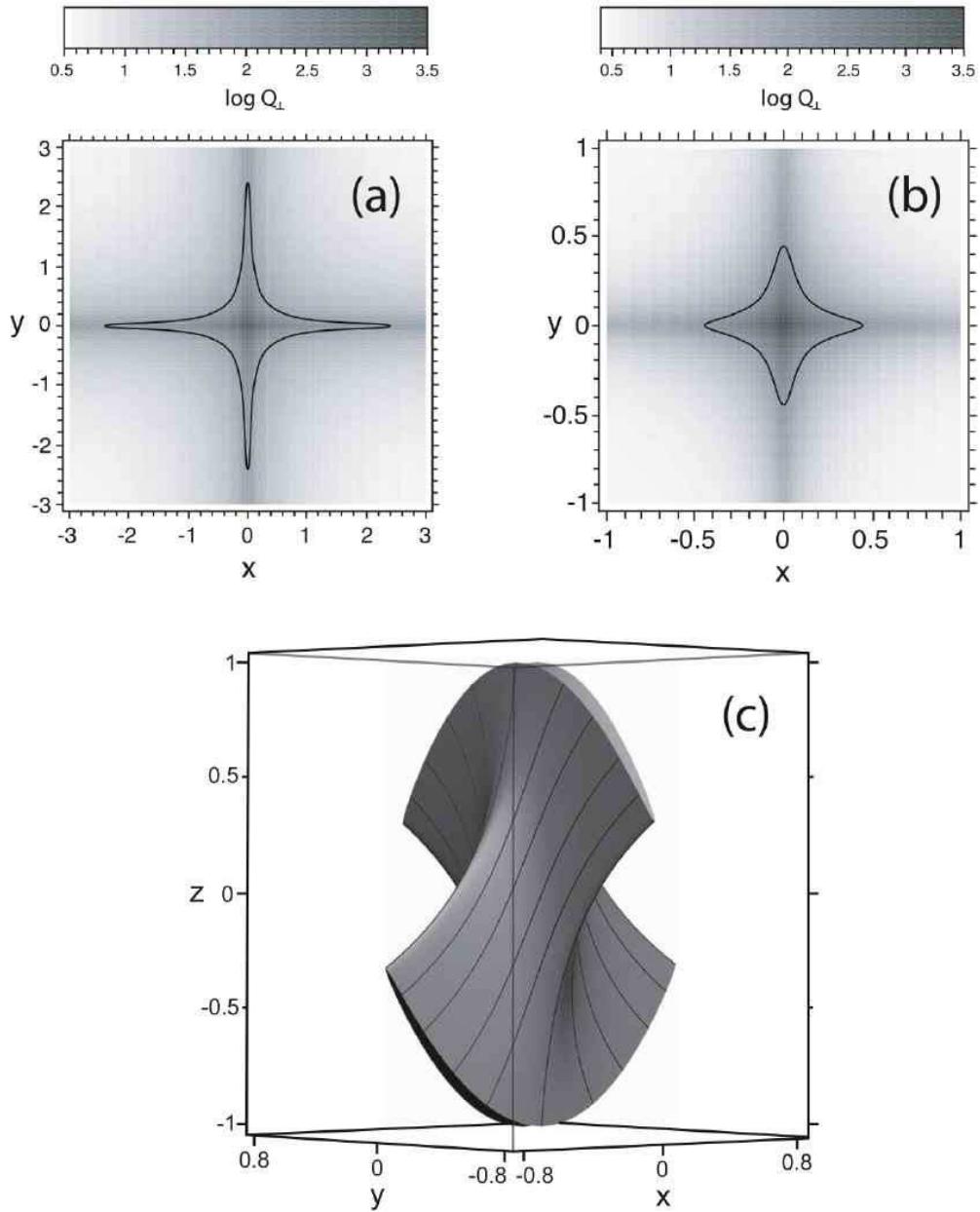}
\caption{Distributions of $\log Q_{\perp}$ in the plane $z=0$ for the X-line configuration from eq. (\ref{BX}) at $h=2$.
The boundaries are two parallel planes $z=\pm1$ ({\it a}) or two hyperbolic paraboloids $h(x^{2}-y^{2})/2-z=\mp1$ ({\it b}).
In the second case, $Q_{\perp}=Q$, since the field lines are strictly orthogonal to the paraboloids.
The contours in both distributions correspond to $Q_{\perp}=100$.
In the second case, such a contour represents also the cross section of the HFT by the plane $z=0$, so that the field lines passing through this contour form the lateral surface of the HFT ({\it c}) bounded from the top and bottom by the hyperbolic paraboloids.
	\label{f:f2} }
\end{figure}

In the particular case of plane boundaries, we have to put $Z_{\pm}(\xi, \eta) = \pm1$ in (\ref{QpX})--(\ref{Ypm}).
Using then (\ref{xi}), (\ref{eta}), (\ref{Xpm}) and (\ref{Ypm}), expression (\ref{QpX}) for $Q_{\perp}$ can be rewritten even as an explicit function of $(x,y,z)$.
By comparing (\ref{QpX}) and (\ref{QX}) one can see also that the terms containing $X_{\pm}$ and $Y_{\pm}$ are responsible in this case for eliminating the projection effect.
As a result of this, the distribution of $Q_{\perp}$ in the $z=0$ plane shows very pronounced ``ridges" along the $x$ and $y$ axes (Fig. \ref{f:f2}a) by revealing an expected HFT with a characteristic X-type intersection of QSLs along the X-line.
According to equations (\ref{x}) and (\ref{y}), such a structure shrinks and expands along these axes exponentially fast with $z$ to give a typical HFT variation of its cross section (see diagram [\ref{-X-}]). 
With growing $h$, the ridges of the $Q_{\perp}$ distribution and so the corresponding QSLs become thinner and thinner by extending on larger and larger distances from the X-line.
Thus, the perpendicular squashing factor $Q_{\perp}$ defines indeed an HFT such that it continuously transforms in the limit of large $h$ into the separatrix planes $x=0$ and $y=0$.

\onecolumn
\twocolumn

However, the peripheral field lines in this HFT are still nearly parallel to the boundaries $z=\pm1$, which causes too strong of a variation of the HFT cross section.
Aesthetically more pleasing HFT in the X-line configuration can be obtained by using the boundaries which are orthogonal to the field lines.
These are iso-surfaces of magnetic potential $F=h\left(x^{2}-y^{2}\right)/2 -z$ having the shape of hyperbolic paraboloids. 
It is natural to chose them passing through the points $(x=y=0,z=\pm 1)$, which means that $F=\mp1$ for such iso-surfaces.
This condition using (\ref{Xpm}) and (\ref{Ypm}) yields 
\begin{eqnarray}
	\frac{h^{2}}{2}\left(\xi^{2} e^{-2hZ_{\pm}} - \eta^{2} e^{2hZ_{\pm}}\right)
	  - Z_{\pm} \pm 1=0 ,
	\label{Zpmeq}
\end{eqnarray}
which is a transcendental equation for the $Z_{\pm}$ functions entering in expression (\ref{QpX}).
This equation is not difficult to solve numerically for given $\xi$ and $\eta$ and to use the respective solution for the calculation of $Q_{\perp}$.
An example of the resulting $Q_{\perp}$ distribution in the plane $z=0$ at $h=2$ is shown in Figure \ref{f:f2}b.
The corresponding HFT with the magnetic surface defined by $Q_{\perp}=100$ and boundaries $F=\mp1$ is presented in Figure \ref{f:f2}c.
For the chosen type of boundaries, $Q_{\perp}=Q$, so both squashing factors define the same HFT.  
The magnetic field for this HFT has the simplest analytical form; the hyperbolic paraboloids are also relatively simple boundary surfaces.
So we believe that this example provides the simplest possible HFT relevant for theoretical studies of basic MHD processes, such as magnetic pinching and reconnection, in three dimensions.


\subsection{HFT in twisted magnetic configuration}
	\label{s:HFTtwst}

The considered X-line configuration with $z=\pm1$ boundaries is an important example, where $Q_{\perp}$, in contrast to $Q$, succeeds in determining an expected HFT.
However, this example is not representative enough to make a general conclusion on the potentialities of $Q$ and $Q_{\perp}$ for detecting QSLs.
Because the field lines in such a configuration behave in a rather artificial way over a major part of the boundaries.
A better comparison of $Q$ and $Q_{\perp}$ can be done by using a more realistic field.

For this purpose, we have chosen the analytical model of a twisted magnetic field \citep{Titov1999}, hereafter called the T\&D model.
It describes approximate equilibria of a circular magnetic flux rope, whose interior force-free field is continuously embedded into a potential background field.
The latter is produced by fictitious subphotospheric sources consisting of two magnetic monopoles of opposite signs and a line current, all located at the axis of symmetry of the rope.
The axis itself is placed some depth below the photospheric plane and the minor radius of the rope $a$ is assumed to be much smaller than the major one $\Rc$ and the distance between the monopoles $L$.

\begin{figure}[htbp]
\epsscale{0.85}
\plotone{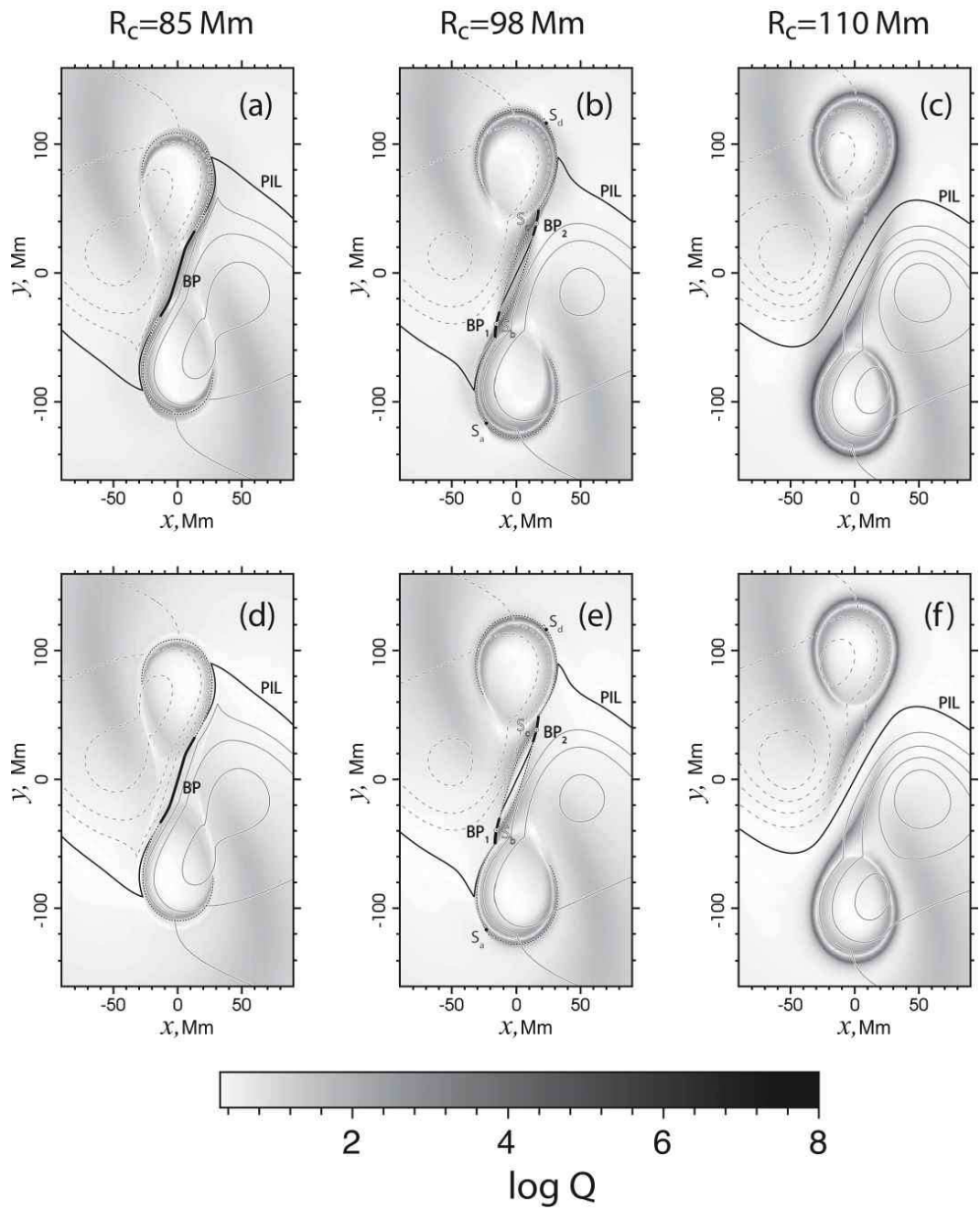}
\caption{Distributions of $\log Q$ ([{\it a\/}], [{\it b\/}] and [{\it c\/}]) vs. $\log Q_{\perp}$ ([{\it d\/}], [{\it e\/}] and [{\it f\/}]) at different values of the major radius $R_{\rm c}$ of the flux rope and with other parameters of the model fixed.
The contours of the photospheric normal magnetic field $B_{z}=0, \pm100, \pm200$ and $\pm400\;{\rm G}$ (and additionally $B_{z}=\pm300\;{\rm G}$ for [{\it c\/}] and [{\it f\/}]), the BPs ({\it thick solid segments of the PIL}), and footprints of the BP separatrices ({\it dotted thin lines}  on [{\it a}], [{\it b}], [{\it d}] and [{\it e}]) are also shown.
The footprints $S_{a}$ and $S_{d}$ and touch points $S_{b}$ and $S_{c}$ of the BP separator field line are indicated on panels ({\it b}) and ({\it e}). 
	\label{f:f3} }
\end{figure}

To compare $Q$ and $Q_{\perp}$ in detail, we have computed their distributions for three sets of parameters which differ only in values of $\Rc$.
Two of these values ($\Rc=85$ and $98\:{\rm Mm}$) and all the remaining parameters are chosen to be exactly the same as in the T\&D model.
By growing $\Rc$ but keeping other parameters fixed we imitate an emergence of the flux rope from below the photosphere.
In this process, the configuration passes continuously through three distinct topological phases.
For sufficiently small $\Rc$, there is a single BP separatrix surface \citep{Titov1999}---the configuration with $\Rc=85\:{\rm Mm}$ represents one of these topological states.
The corresponding distributions of $Q$ and $Q_{\perp}$ are shown in Figures \ref{f:f3}a and \ref{f:f3}d, respectively.
With growing $\Rc$, this BP and the associated separatrix surface bifurcate into two parts to give birth to a BP separator field line \citep{Titov1999}---the configuration with $\Rc=98\:{\rm Mm}$ represents this second topological phase.
The corresponding distributions of $Q$ and $Q_{\perp}$ are shown in Figures \ref{f:f3}b and \ref{f:f3}e, respectively.
The points ${\rm S}_{\rm a}$ and ${\rm S}_{\rm d}$ on these figures are the footpoints of the BP separator, while ${\rm S}_{\rm b}$ and ${\rm S}_{\rm c}$ are its photospheric contact points.

Further growing of $\Rc$ leads to a complete disappearance of the bifurcated BP and the associated separatrix surfaces---the configuration with $\Rc=110\:{\rm Mm}$ represents this last phase.
The corresponding distributions of $Q$ and $Q_{\perp}$ are shown in Figures \ref{f:f3}c and \ref{f:f3}f, respectively.
The magnetic field at these parameters becomes topologically trivial, since its field line mapping is continuous everywhere.
The whole structure can be continuously transformed to a simple arcade-like configuration with the help of a suitable photospheric motion.
Thus, with growing $\Rc$ or emerging of the flux rope
\onecolumn
\begin{figure}[htbp]
\epsscale{0.85}
\plotone{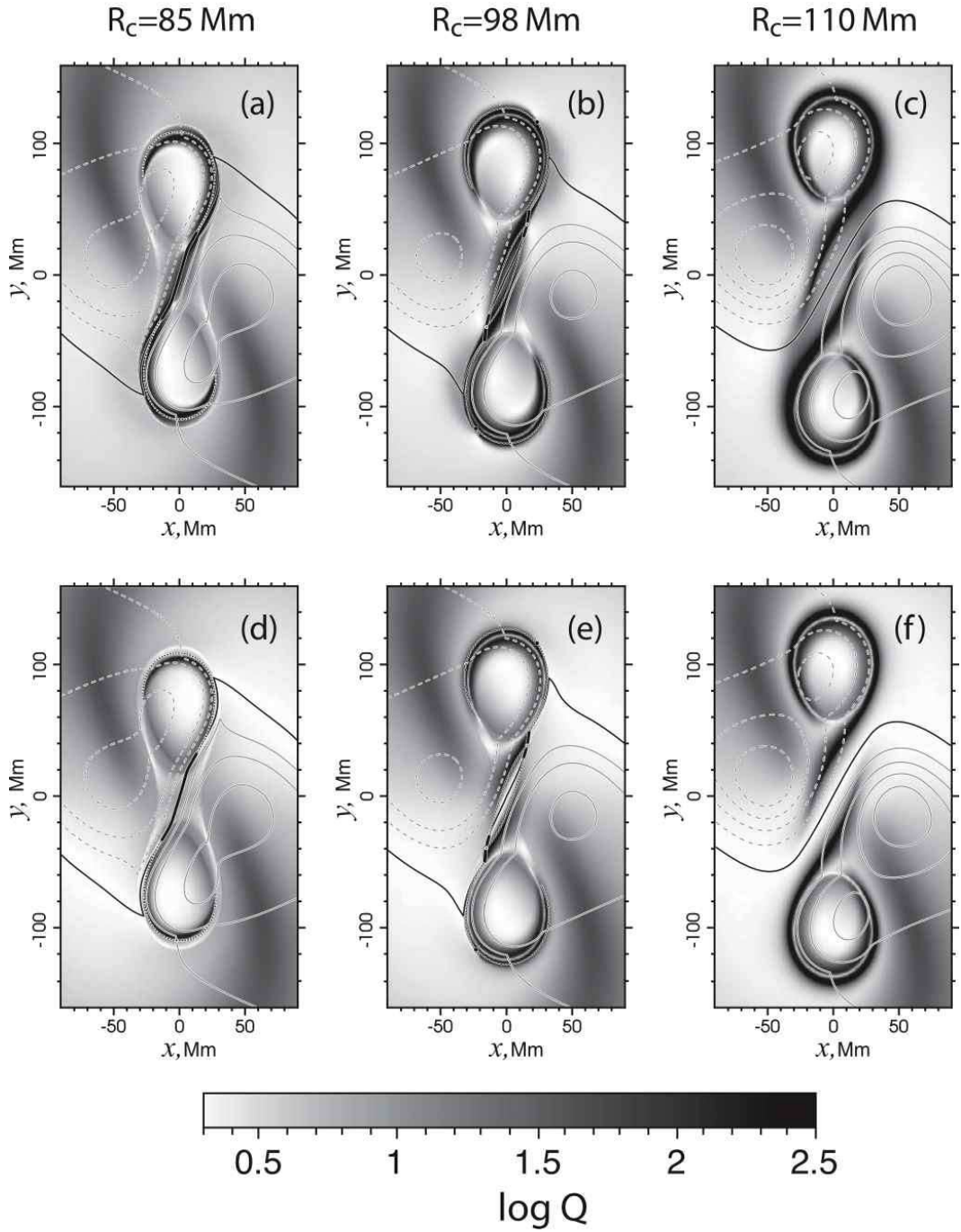}
\caption{Same as Fig. \ref{f:f3}, but with the shading saturated at the level $2.5$ to reveal the difference between the distributions at moderate values.
	\label{f:f4} }
\end{figure}
\twocolumn
{ \parindent=0pt
the topological complexity of the configuration first increases and then abruptly decreases.}

On the contrary, the geometrical distortion of the field line mapping gradually increases in the configuration during this process.
Both squashing factors $Q$ and $Q_{\perp}$ continue to grow with $\Rc$ in narrow strips of the photospheric plane.
It is clearly seen from the $Q$ and $Q_{\perp}$ distributions plotted with the same grayscaling (Fig. \ref{f:f3}) that $Q > Q_{\perp}$ everywhere and the difference between $Q$ and $Q_{\perp}$ becomes smaller with growing $\Rc$.
The most significant difference between them is seen at the first two phases  near the BPs and the footprints of the associated separatrix surfaces.
As previously anticipated, $Q$ always rises in these regions, while $Q_{\perp}$ does not, except near the contact points ${\rm S}_{\rm b}$ and ${\rm S}_{\rm c}$ of the BP separator (Fig. \ref{f:f3}e).
The value $Q_{\perp}$ does rise there but only to give birth to a part of HFT footprints, which are matured eventually in the third phase (Fig. \ref{f:f3}f).
As concerned with the indicated rise of $Q$, it is mainly caused by the projection effect: the field lines which are close to the BP separatrix surfaces approach the photosphere near the BPs at a small angle to the horizontal, which strongly distorts the footprints of the corresponding elemental flux tubes.
Comparing $Q$ and $Q_{\perp}$ at $\Rc=110\:{\rm Mm}$ reveals that the latter is valid for the central part of the PIL as well.

\begin{figure}[htbp]
\epsscale{1.0}
\plotone{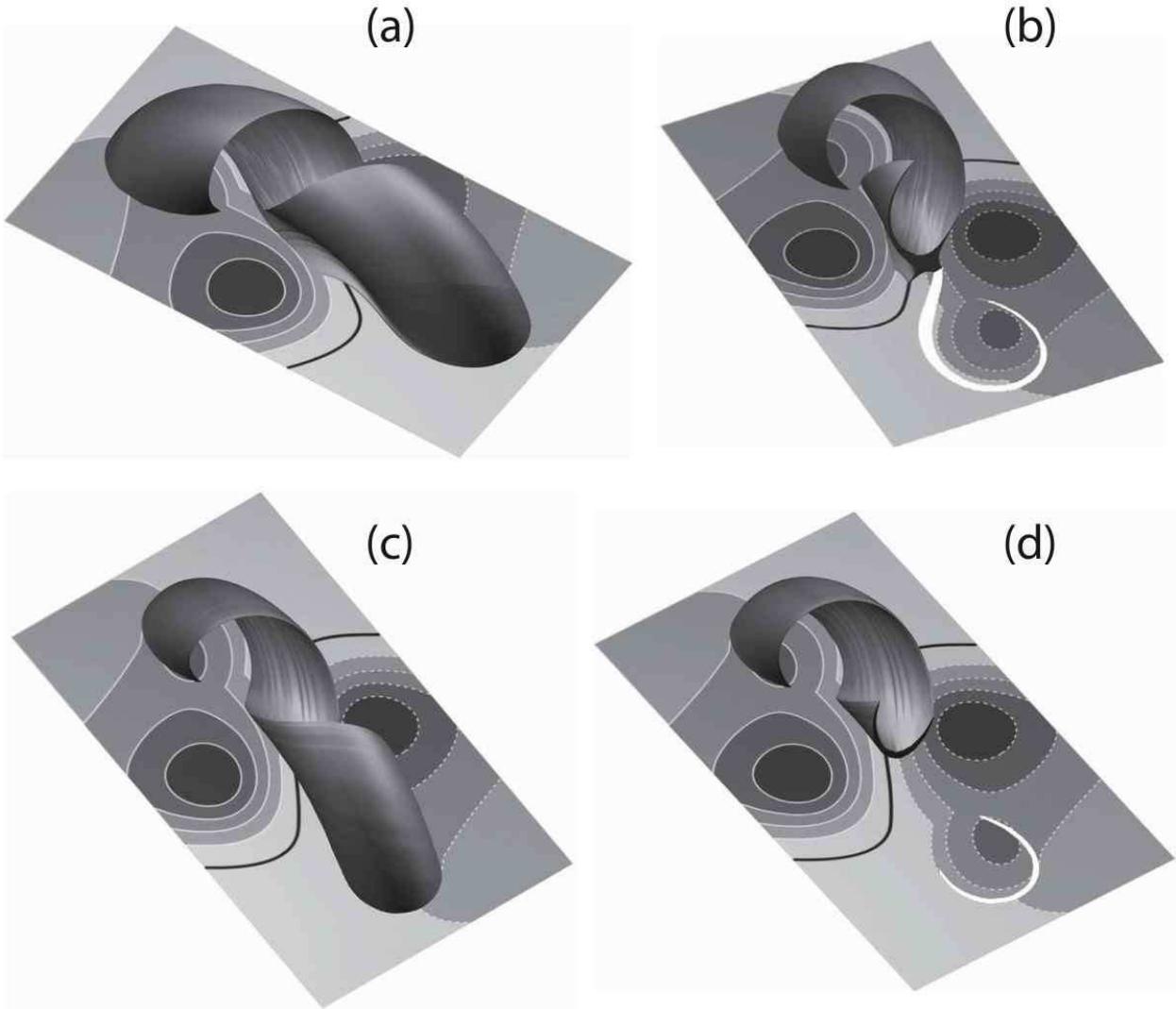}
\caption{HFT ({\it a}), the helical QSL ({\it c}) and their cross sections, for the HFT ({\it b}) and for the QSL ({\it d}), in the T\&D twisted magnetic configuration at $R_{\rm c}=110\;{\rm Mm}$.
Other parameters of the model, as well as the contours of the photospheric normal magnetic field $B_{z}$, are the same as in Figs. \ref{f:f3}c and \ref{f:f3}f; the white stripes on the photosphere represent the HFT and QSL footprints, whose contours correspond to $Q=100$.
	\label{f:f5} }
\end{figure}

The discussed features of the distributions become more transparent if one saturates the grey shading in the plots at the values $\ge 2.5$ (Fig. \ref{f:f4}).
These new plots show that the QSLs based on the $Q_{\perp}$ distribution are characterized by a thinner and more uniform thickness.
Their footprints acquire at $\Rc=110\:{\rm Mm}$ a clear fishhook-like shape in each of the photospheric polarities with maximums of $Q$ reaching $\sim 10^{8}$.
The QSLs rooted at such ``fishhooks" intersect each other by combining themselves into an HFT \citep{Titov2003a}.
This structural feature seems to be very robust, because it appears even in twisted configurations which are not in force-free or magnetostatic equilibrium \citep{Demoulin1996b}.
One can see from Figure \ref{f:f5}a that, except for an essential twisting distortion, the cross section of such an HFT varies exactly according to diagram (\ref{-X-}).

It has yet to be proved, but it seems to be quite natural that this HFT is pinched into a vertical current sheet below the flux rope by its upward movement when the kink or torus instability is developed in the configuration at a sufficiently large twist of the field lines in the rope \citep{Toeroek2004, Roussev2003}.
This interpretation is very important for understanding the properties of sigmoidal structures in flaring configurations \citep{Kliem2004}.
Figure \ref{f:f5}a suggests that the sigmoids in such configurations are simply pinching HFTs illuminated by a hot plasma material which appears there due to the reconnection process in the above-mentioned vertical current sheet.
This seems to be valid at least for the third topological phase of the flux rope emergence, while an additional interaction with the photosphere must be involved at the first and second topological phases, where the BPs are present \citep{Titov1999, Fan2003}.
Panels ({\it d})---({\it f}) in Figures \ref{f:f3} and \ref{f:f4} demonstrate that the footprints of the BP separatrix surfaces follow very close to the HFT footprints emerging gradually with growing $\Rc$.
This implies the corresponding similarity in the shapes of such separatrix surfaces and HFTs.
So the explanations of the sigmoids that rely on either the presence of the BPs \citep{Titov1999, Fan2003} or the HFTs \citep{Kliem2004} are not alternative but rather complementary, since they refer to different phases of the flux rope emergence.

It should be noted also that both $Q$ and $Q_{\perp}$ distributions (Fig. \ref{f:f4}) contain at the border of the flux rope two less pronounced horseshoe-like features with maximums $\sim 10^{2}$.
The QSL rooted at these ``horseshoes" has a helical shape (Fig. \ref{f:f5}b) with a slightly varying cross section along the field lines.
So this QSL has a structure qualitatively different from those two which form the HFT.
The comparison of $Q$ and $Q_{\perp}$ shows that the squashing of the flux tubes in this helical QSL is only in part due to the projection effect.
The $Q_{\perp}$ distribution demonstrates that the major contribution to $Q$ comes from the shearing of the twisted field lines in the rope.
Thus, both distributions reveal a helical QSL which is a part of the inner border layer of the flux rope.
In this respect, the considered example demonstrates that our squashing factors help identify flux ropes themselves.
It is not a problem, of course, to locate the flux rope in the T\&D model, where its parameters are known from the construction of the model.
Yet identifying flux ropes in more complicated configurations obtained, for instance, numerically from magne-
\onecolumn
\twocolumn
{ \parindent=0pt
togram data is a real problem.
If such configurations are topologically trivial, like those in the third phase of our example, the determination of QSLs seems to be the only method for identifying flux ropes.}

As shown above, both squashing factors allow us in the case of the T\&D model to determine similar QSLs, except that $Q_{\perp}$ is more advanced than $Q$ near BPs, whenever they appear. 
Therefore, we think that, in general, if the numerical grid used for computing $Q_{\perp}$ is fine enough for detecting possible BP separatrix surfaces by sudden spikes in the $Q_{\perp}$ distribution, the use of only $Q_{\perp}$ would be sufficient for the structural analysis of configurations. 
In practice, however, the required resolution of the grid cannot be always easily foreseen.
Also the computational cost for $Q$ is not really high, since the same input data as for $Q_{\perp}$ can be used. 
So it is sensible to compute both these distributions at a time and compare them in the same way as we did in the considered example.
Some redundancy of the information contained in such distributions is not superfluous but useful, especially, in the case of complicated real magnetic configurations.
Thus, from the practical point of view, the value $Q$ should not be considered as obsolete but rather as a complementary characteristic of magnetic connectivity.


\section{SUMMARY}
	\label{s:s}

We have derived a covariant form of the squashing factor $Q$, which enables us to determine quasi-separatrix layers (QSLs) in both closed and open magnetic configurations with an arbitrary shape of boundaries.
The corresponding expression for $Q$ assumes that the Jacobian matrix of the field line mapping and the metric tensors at the footpoints are known.
The expression admits also that such ``input data" can be represented with the help of two different coordinate systems for determining location of the conjugate footpoints on the boundaries.
This provides a firm theoretical basis for a global description of the field line connectivity in the solar corona.

To eliminate the projection effect at the field lines which are nearly touching the boundary, the perpendicular squashing factor $Q_{\perp}$ is also derived in a similar covariant form.
The value $Q_{\perp}$ defines the degree of squashing of elemental magnetic flux tubes only in the directions orthogonal to the field lines.
In the definition of $Q_{\perp}$, the boundaries enclosing the magnetic configuration constrain only the length of the flux tubes while not affecting their cross sections at the footpoints.
For calculating $Q_{\perp}$, the vectors of magnetic field at the footpoints  are required in addition to the same input data as for $Q$.
The use of both covariant squashing factors is demonstrated by calculating them for the boundaries with the planar and spherical geometries.
Then the properties of $Q$ and $Q_{\perp}$ are compared by considering two examples of magnetic configurations.

The first example is a classical X-line configuration of potential magnetic field in a plasma volume restricted by two boundary surfaces.
It is easy to show that for the plane boundaries perpendicular to the X-line, the value $Q$ is constant.
So the $Q$ distribution does not allow us to define any QSLs in such a configuration.
The reason for this failure lies in the projection effect, which is very strong for the field lines distant from the X-line. 
We have also calculated an analytical expression of $Q_{\perp}$ for the same field but with the boundaries of an arbitrary shape.
In the case of the plane boundaries, this new value $Q_{\perp}$, in contrast to $Q$, has a non-uniform distribution, which does reveal the expected two QSLs.
These QSLs intersect each other by combining themselves into a hyperbolic flux tube (HFT).
A more elegant HFT is obtained for the X-line configuration with the boundaries orthogonal to the field lines.

To make a better comparison of the properties of $Q$ and $Q_{\perp}$, a second magnetic configuration more relevant for solar physics is considered.
The respective field is defined by using the Titov-D\'{e}moulin (1999) model of a force-free flux rope embedded into a potential background field.
Contrary to the case of the X-line configuration restricted by the plane boundaries, both the $Q$ and $Q_{\perp}$ distributions reveal QSLs in the twisted configuration.
These distributions are similar everywhere except near bald patches (BPs) and footprints of the associated separatrix surfaces, whenever the BPs exist.
By definition, the value $Q_{\perp}$ is free of the projection effect, so that $Q_{\perp}$ rises near BPs only if the corresponding flux tubes are subject to a volumetric squashing.
This is not the case, of course, for the value $Q$, which always rises in such regions of the photosphere.
So, in comparison with $Q$, the value $Q_{\perp}$ shows itself to be again a superior characteristic for analysis of magnetic connectivity.

Nevertheless, we have argued that it is more practical in general to compute both squashing factors for analyzing the structure of a given magnetic configuration.
This does not require additional significant effort, while making it easy to discriminate between the volumetric squashing of elemental flux tubes and the surface projection effect at the boundaries.

\acknowledgments

This research was supported by NASA and the Center for Integrated
Space Weather Modeling (an NSF Science and Technology Center).




\begin{thebibliography}{28}
\expandafter\ifx\csname natexlab\endcsname\relax\def\natexlab#1{#1}\fi

\bibitem[{{Arnold}(1988)}]{Arnold88}
{Arnold}, V.~I. 1988, {Geometrical methods in the theory of ordinary
  differential equations} (New York, Springer-Verlag, 364 p.)

\bibitem[{{Baum} \& {Bratenahl}(1980)}]{Baum1980}
{Baum}, P.~J., \& {Bratenahl}, A. 1980, \solphys, 67, 245

\bibitem[{{D\'{e}moulin} {et~al.}(1997){D\'{e}moulin}, {Bagala}, {Mandrini},
  {Henoux}, \& {Rovira}}]{Demoulin1997}
{D\'{e}moulin}, P., {Bagala}, L.~G., {Mandrini}, C.~H., {Henoux}, J.~C., \&
  {Rovira}, M.~G. 1997, \aap, 325, 305

\bibitem[{{D\'{e}moulin} {et~al.}(1996a){D\'{e}moulin}, {Henoux}, {Priest}, \&
  {Mandrini}}]{Demoulin1996a}
{D\'{e}moulin}, P., {Henoux}, J.~C., {Priest}, E.~R., \& {Mandrini}, C.~H.
  1996a, \aap, 308, 643

\bibitem[{{D{\'e}moulin} {et~al.}(1996b){D{\'e}moulin}, {Priest}, \&
  {Lonie}}]{Demoulin1996b}
{D{\'e}moulin}, P., {Priest}, E.~R., \& {Lonie}, D.~P. 1996b, \jgr, 101, 7631

\bibitem[{{Fan} \& {Gibson}(2003)}]{Fan2003}
{Fan}, Y., \& {Gibson}, S.~E. 2003, \apjl, 589, L105

\bibitem[{{Kageyama} \& {Sato}(2004)}]{Kageyama2004}
{Kageyama}, A., \& {Sato}, T. 2004, Geochemistry, Geophysics, Geosystems, 5,
  9005

\bibitem[{{Kliem} {et~al.}(2004){Kliem}, {Titov}, \&
  {T{\"o}r{\"o}k}}]{Kliem2004}
{Kliem}, B., {Titov}, V.~S., \& {T{\"o}r{\"o}k}, T. 2004, \aap, 413, L23

\bibitem[{{Lau} \& {Finn}(1990)}]{Lau1990}
{Lau}, Y.-T., \& {Finn}, J.~M. 1990, \apj, 350, 672

\bibitem[{{Levine} {et~al.}(1982){Levine}, {Schulz}, \& {Frazier}}]{Levin1982}
{Levine}, R.~H., {Schulz}, M., \& {Frazier}, E.~N. 1982, \solphys, 77, 363

\bibitem[{{Longcope}(2001)}]{Longcope2001}
{Longcope}, D.~W. 2001, Physics of Plasmas, 8, 5277

\bibitem[{{Longcope} \& {Cowley}(1996)}]{Longcope1996}
{Longcope}, D.~W., \& {Cowley}, S.~C. 1996, Physics of Plasmas, 3, 2885

\bibitem[{{Longcope} \& {Strauss}(1994{\natexlab{a}})}]{Longcope1994a}
{Longcope}, D.~W., \& {Strauss}, H.~R. 1994{\natexlab{a}}, \apj, 426, 742

\bibitem[{{Longcope} \& {Strauss}(1994{\natexlab{b}})}]{Longcope1994b}
---. 1994{\natexlab{b}}, \apj, 437, 851

\bibitem[{{Marsden} {et~al.}(2002){Marsden}, {Ratiu}, \&
  {Abraham}}]{Marsden2002}
{Marsden}, J.~E., {Ratiu}, T., \& {Abraham}, R. 2002, {Manifolds, Tensor
  Analysis, and Applications}, 3rd edn., Applied Mathematical Sciences (New
  York, Springer-Verlag, 555 p.)

\bibitem[{{Parker}(1979)}]{Parker1979}
{Parker}, E.~N. 1979, {Cosmical magnetic fields: Their origin and their
  activity} (Oxford, Clarendon Press; New York, Oxford University Press, 858
  p.)

\bibitem[{{Parker}(1994)}]{Parker1994}
---. 1994, {Spontaneous current sheets in magnetic fields: with applications
  to stellar x-rays} (New York, Oxford University Press)

\bibitem[{{Priest} \& {D{\'e}moulin}(1995)}]{Priest1995}
{Priest}, E.~R., \& {D{\'e}moulin}, P. 1995, \jgr, 100, 23443

\bibitem[{{Priest} \& {Forbes}(2000)}]{Priest2000}
{Priest}, E.~R., \& {Forbes}, T.~G. 2000, {Magnetic Reconnection: MHD theory
  and applications} (Cambridge, UK: Cambridge University Press, 612 p.)

\bibitem[{{Priest} \& {Titov}(1996)}]{Priest1996}
{Priest}, E.~R., \& {Titov}, V.~S. 1996, Philos. Trans. R. Soc. London A, 354,
  2951

\bibitem[{{Roussev} {et~al.}(2003){Roussev}, {Forbes}, {Gombosi}, {Sokolov},
  {DeZeeuw}, \& {Birn}}]{Roussev2003}
{Roussev}, I.~I., {Forbes}, T.~G., {Gombosi}, T.~I., {Sokolov}, I.~V.,
  {DeZeeuw}, D.~L., \& {Birn}, J. 2003, \apjl, 588, L45

\bibitem[{{Sweet}(1969)}]{Sweet1969}
{Sweet}, P.~A. 1969, \araa, 7, 149

\bibitem[{{Syrovatskii}(1981)}]{Syrovatskii1981}
{Syrovatskii}, S.~I. 1981, \araa, 19, 163

\bibitem[{{Titov} \& {D{\'e}moulin}(1999)}]{Titov1999}
{Titov}, V.~S., \& {D{\'e}moulin}, P. 1999, \aap, 351, 707

\bibitem[{{Titov} {et~al.}(1999){Titov}, {D{\'e}moulin}, \&
  {Hornig}}]{Titov1999a}
{Titov}, V.~S., {D{\'e}moulin}, P., \& {Hornig}, G. 1999, in ESA SP-448:
  Magnetic Fields and Solar Processes, ed. A.~{Wilson} \& {et al.}, 715--722

\bibitem[{{Titov} {et~al.}(2003){Titov}, {D\'{e}moulin}, \&
  {Hornig}}]{Titov2003a}
{Titov}, V.~S., {D\'{e}moulin}, P., \& {Hornig}, G. 2003, Astronomische
  Nachrichten, 324, 17

\bibitem[{{Titov} \& {Hornig}(2002)}]{Titov2002a}
{Titov}, V.~S., \& {Hornig}, G. 2002, Advances in Space Research, 29, 1087

\bibitem[{{Titov} {et~al.}(2002){Titov}, {Hornig}, \&
  {D{\'e}moulin}}]{Titov2002}
{Titov}, V.~S., {Hornig}, G., \& {D{\'e}moulin}, P. 2002, \jgr, 107, 3

\bibitem[{{Titov} {et~al.}(1993){Titov}, {Priest}, \& {Demoulin}}]{Titov1993}
{Titov}, V.~S., {Priest}, E.~R., \& {Demoulin}, P. 1993, \aap, 276, 564

\bibitem[{{T{\"o}r{\"o}k} {et~al.}(2004){T{\"o}r{\"o}k}, {Kliem}, \&
  {Titov}}]{Toeroek2004}
{T{\"o}r{\"o}k}, T., {Kliem}, B., \& {Titov}, V.~S. 2004, \aap, 413, L27

\end{thebibliography}



\end{document}